\documentclass[letterpaper]{article} 
\usepackage{aaai2027}  

\nocopyright
\usepackage[hyphens]{url}  
\usepackage{graphicx} 
\usepackage{natbib}  
\usepackage{caption} 
\usepackage{algorithm}
\usepackage{algorithmic}

\usepackage{amsmath}
\usepackage{amssymb}
\usepackage{multirow}
\usepackage{subcaption}

\usepackage{xcolor}

\usepackage{newfloat}
\usepackage{listings}
\DeclareCaptionStyle{ruled}{labelfont=normalfont,labelsep=colon,strut=off} 
\floatstyle{ruled}
\newfloat{listing}{tb}{lst}{}
\floatname{listing}{Listing}

\usepackage{booktabs}

\title{Text2Score: Generating Sheet Music From Textual Prompts}
\author {
    Keshav Bhandari\textsuperscript{\rm 1}\corresponding,
    Sungkyun Chang\textsuperscript{\rm 1,\rm 4},
    Abhinaba Roy\textsuperscript{\rm 2},
    Francesca Ronchini\textsuperscript{\rm 3},
    Emmanouil Benetos\textsuperscript{\rm 1},
    Dorien Herremans\textsuperscript{\rm 2},
    Simon Colton\textsuperscript{\rm 1}
}
\affiliations {
    \textsuperscript{\rm 1}Queen Mary University of London, UK\\
    \textsuperscript{\rm 2}Singapore University of Technology and Design, Singapore\\
    \textsuperscript{\rm 3}Politecnico di Milano, Italy\\
    \textsuperscript{\rm 4}EmotionWave, Korea\\
    k.bhandari@qmul.ac.uk
}

\begin{document}

\maketitle

\begin{abstract}
Developing text-driven symbolic music generation models remains challenging due to the scarcity of aligned text-music datasets and the unreliability of automated captioning pipelines. While most efforts have focused on MIDI, sheet music representations are largely underexplored in text-driven generation. We present \textit{Text2Score}, a two-stage framework comprising a planning stage and an execution stage for generating sheet music from natural language prompts. By deriving supervision signals directly from symbolic XML data, we propose an alternative to caption-based training that bypasses noisy or scarce text-music pairs. In the planning stage, an LLM orchestrator translates a natural language prompt into a structured bar-wise plan defining musical attributes such as instruments, key, time signatures, harmony, etc. This plan guides a generative model in the execution stage to produce interleaved ABC notation conditioned on its structural constraints. To assess output quality, we introduce an evaluation framework covering playability, readability, instrument utilization, structural complexity, and prompt adherence, corroborated by expert musicians. \textit{Text2Score} consistently outperforms both a pure LLM-based agentic framework and three end-to-end baselines across objective and subjective dimensions. We open-source the dataset, code, evaluation set and LLM prompts used in this work\footnote{https://github.com/keshavbhandari/text2score/}; a demo is available on our project page\footnote{https://keshavbhandari.github.io/portfolio/text2score}.
\end{abstract}


\section{Introduction}\label{sec:introduction}
Textual descriptions have become a popular way to guide the generation of symbolic music. This modality typically represents music as either performance-based MIDI signals or notation-based sheet music, such as MusicXML and ABC notation \cite{le:tel-05426752}. While performance MIDI captures expressive signals, sheet music is valued by composers and musicians for its structured arrangements and precise formatting for composition, performance and analysis.

Despite recent progress, developing models that accurately follow text prompts remains difficult due to the scarcity of high-quality, large-scale datasets that pair music with natural language \cite{xu2024generating,bhandari2024motifs}. Many current approaches rely on datasets with features extracted by probabilistic models and subsequent automated captioning with Large Language Models (LLMs). These methods may face data alignment issues \cite{li2026midilm} and LLM hallucinations \cite{doh2023lp}, yielding unreliable text-music descriptions. Furthermore, most current text-to-token models are trained on paired music-caption data in an end-to-end fashion. As available datasets do not support the training of intermediate logic, such models often lack the reasoning capabilities necessary to handle complex musical structures. Evaluation compounds this gap as prior work assesses semantic and metadata alignment, but not whether the resulting score can actually be read and played.

To address these issues, we present \textit{Text2Score}, a two-stage framework that utilizes sub-task decomposition to separate the generation process into a planning stage and an execution stage. In the planning stage, an LLM acts as an orchestrator to translate a user's textual prompt into a structured bar-wise plan. This decomposition provides a specific scope for musical reasoning, allowing the LLM to determine structural elements such as instruments, key and time signature, pitch range, note density, chord note pitches and dynamics before any notes are generated. In the execution stage, this LLM generated plan is fed into a generative model that is trained from scratch with the same musical features (or training plan) extracted directly from the XML training data. We extend the hierarchical decoder architecture of NotaGen \cite{wang2025notagen} with a BERT-based encoder to process the bar-wise plan via cross-attention. As the plan is derived directly from the source music, generation is grounded in the intended musical structure rather than inferred from noisy text-music pairs. We summarize our contributions as follows:

\begin{enumerate}
    \item We introduce \textit{Text2Score}, a two-stage framework pairing an LLM orchestrator for structural planning with a plan-conditioned autoregressive decoder for execution to bridge textual prompts and sheet music generation.
    \item We present an evaluation framework designed to quantify the readability and playability of generated scores, which is further corroborated by expert musicians.
    \item We release the ABC notation dataset used in this work strictly for non-commercial research purposes to support further studies in symbolic sheet music generation.
\end{enumerate}

\begin{figure*}
  \centering
  \includegraphics[alt={AAAI 2027 template example image},width=1.0\linewidth]{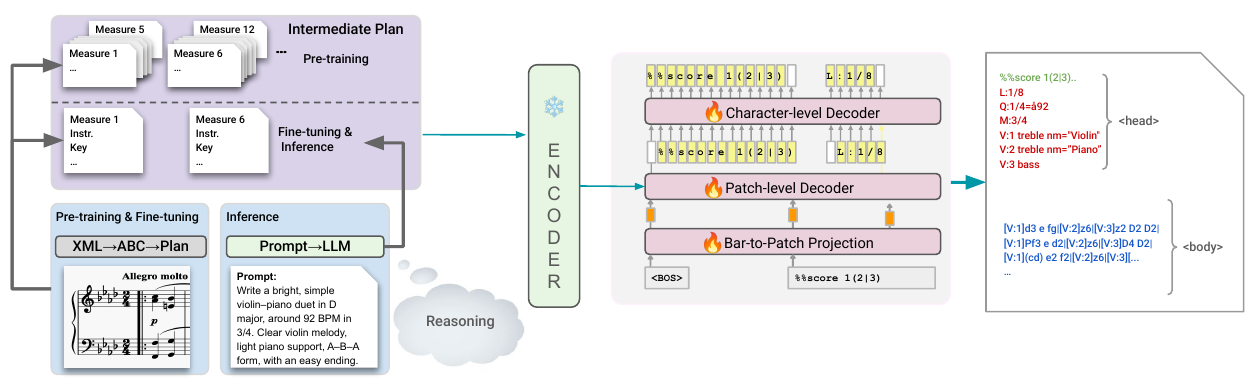}
  \caption{\textit{Text2Score} framework. During pre-training, consecutive bar-wise plans are extracted from symbolic XML; fine-tuning uses a sparse subset of structurally significant pivot measures. At inference, an LLM orchestrator translates a natural language prompt into a structured plan, encoded by ModernBERT and provided to the patch-level decoder via cross-attention. Finally, the character-level decoder takes the patch decoder hidden states to produce interleaved ABC notation.}
  \label{fig:system_overview}
\end{figure*}

\section{Related Work}

\paragraph{Text-to-Symbolic Music Generation:} Early works in text-controlled music generation focused on bridging semantic embeddings and musical representations. Butter \cite{zhang2020butter} aligned sentences with musical sequences via a cross-modal VAE latent space, while \cite{lu2023musecoco} predicted intermediate attributes from text to condition token decoding.

Recent advancements have shifted toward end-to-end training paradigms. Text2midi \cite{bhandari2025text2midi} and Text2midi-InferAlign \cite{roy2025text2midi} pair a text encoder with an autoregressive decoder, while \cite{wu2025midi, li2026midilm} adapted LLM architectures to treat MIDI as a native tokenized language. The authors of \cite{xu2024generating} use LLM-enhanced datasets for richer supervision. However, we hypothesize that these end-to-end systems exhibit poor text adherence and struggle with longer, structured generation due to the absence of a clear intermediate reasoning and planning stage.

Alternative strategies propose other means of control: \cite{wang2024melotrans} applies motif development rules, while \cite{tian2025xmusic} supports multiple input modalities with emotional control. Both require extensive pre-training on large-scale paired datasets. Closest to our conditioning scheme, FIGARO \cite{von2022figaro} generates MIDI from fine-grained, bar-level descriptions of time signature, note density, pitch, velocity, instruments and chords, but relies on users during inference to edit these features directly with no natural language interface. 

\paragraph{LLM-Based Agentic Composition:} A burgeoning area of research investigates the ``musical world'' knowledge implicitly held by LLMs trained solely on text. As shown in \cite{shin2025large}, text-only LLMs can infer rudimentary musical structures and temporal relationships from string-based patterns without explicit musical training. This internal perception facilitates agentic frameworks as seen in ComposerX \cite{deng2024composerx} and CoComposer \cite{xing2025cocomposer} that use LLMs as zero-shot composers. However, LLMs acting as the sole generative engine often produce syntactically inconsistent or musically simplistic outputs. \textit{Text2Score} occupies a middle ground, leveraging LLM reasoning for structural planning while delegating score execution to a dedicated model.

\paragraph{Structural Planning and Hierarchical Architectures:} Using LLMs to decipher prompts into plans was explored in $M^6(GPT)^3$ \cite{pocwiardowski2025m6}, which initialized genetic algorithms for melody generation. Our execution stage adopts a hierarchical decoder, a design choice well supported by prior hierarchical architectures and representations in symbolic music generation \cite{wu2019hierarchical, 10089423, zixun2021hierarchical, dai2021controllable, zhang2022structure, wang2025notagen}. Among these, NotaGen \cite{wang2025notagen} introduced bar level hierarchy, a natural choice given the metrical structure of music and the bar-wise organisation of interleaved ABC notation, where all voices are consolidated into a single line per bar compared to that of vanilla ABC notation. This granularity aligns directly with our bar-wise plan. Unlike NotaGen, which is limited to ``period-composer-instrumentation'' prompts, our framework accepts free-form natural language.

ABC notation as a text-based sheet music representation has grown popular for its near-lossless XML conversion as seen in studies such as \cite{sturm2015folk, wu2023tunesformer, qu2024mupt, zhou2025emelodygen, liang2024bytecomposer, wu2024melodyt5, wang2025notagen, kumar2026farpretrainedllmssymbolic}. While these works demonstrate the efficiency of ABC notation for composition modelling, our work also evaluates the readability and playability of the generated sheet music.

\section{Methods}
\label{sec:methods}

As shown in Figure \ref{fig:system_overview}, we first describe the plan structure during training, before introducing our generative model's architecture that uses it in the execution stage.

\subsection{The Structural Plan}
We formally define the bar-wise plan $\mathcal{P}$ as a structured skeleton of the music, comprising a sequence of structural descriptors derived directly from the symbolic XML source. The plan is defined as $\mathcal{P} = \{N, G, I_{total}, \mathbf{m}_1, \mathbf{m}_2, \dots, \mathbf{m}_N\}$, where $N$ is the total number of bars, $G$ represents the piece's genre if available and $I_{total}$ is the complete instrumentation set. Each bar-specific vector $\mathbf{m}_i$ is represented as:
\begin{equation}
    \mathbf{m}_i = \{I_i, R_i, D_i, T_i, \theta_i, \kappa_i, C_i, Y_i\}
\end{equation}
where $I_i$ is the set of active instruments, $R_i$ is the pitch range (MIDI min/max), $D_i$ is the categorical note density (low, medium, high), $T_i$ is the tempo, $\theta_i$ and $\kappa_i$ are the time and key signatures, $C_i$ is the pitch-class set representing the harmony, and $Y_i$ represents the expressive dynamics for bar $i$ if it exists.

\subsection{Model Architecture}

Our execution model pairs an encoder with an autoregressive decoder to generate interleaved ABC. We emphasize that our framework is agnostic to the execution-stage backbone: any autoregressive transformer conditioned on the plan via cross-attention may suffice. As our contribution lies in the plan-conditioned training paradigm rather than a bespoke architecture, we adopt NotaGen's \cite{wang2025notagen} hierarchical decoder principally for its bar-level granularity, which aligns naturally with our bar-wise plan.

\textbf{Plan Encoder:} We use ModernBERT \cite{warner2025smarter} as our frozen encoder to process the plan $\mathcal{P}$. By encoding the plan, the model extracts contextualized latent representations $H_{plan} \in \mathbb{R}^{N \times d}$, which serve as the conditioning for the generation process.

\textbf{Patch-level Decoder:} Following the bar-patching scheme of NotaGen, each bar of interleaved ABC is treated as a single \textit{patch}: a fixed-length block of characters that the Bar-to-Patch Projection (Fig. \ref{fig:system_overview}) compresses into one embedding, so the decoder operates over a sequence of bars compared to raw characters. A GPT-based model captures temporal dependencies across these bar patches. At each patch position $j$, it applies causal self-attention over preceding patches together with cross-attention $\text{Attn}(Q, K, V)$, where the queries $Q$ are patch representations and the keys/values $K, V$ derive from $H_{plan}$, grounding each bar in the plan's constraints. The decoder's output at position $j$ is the patch hidden state $h_{patch, j}$ which summarizes the intended content of bar $j$.

\textbf{Character-level Decoder:} A lightweight character-level decoder auto-regressively predicts interleaved ABC notation characters for each patch $j$, where the probability of character $c_{j,k}$ is conditioned on the patch hidden state and preceding characters $P(c_{j,k} \mid c_{j,<k}, h_{patch,j})$. The patch hidden state guides the generation of individual musical elements such as specific notes and rhythms to match the structural requirements of the designated bar.

\subsection{Training and Generation}

We employ a two-stage training strategy to promote adherence to the plan while retaining the model’s capacity for autonomous generation.

\textbf{Sequential Pre-training:} The model is first pre-trained on the ABC dataset using consecutive bars of the plan $\mathcal{P}$. This allows the decoder to learn the fundamental mapping between structural descriptors and their corresponding character-level notation.

\textbf{Structural Fine-tuning:} To minimize the gap between our training plans and the plans generated by the LLM during inference, we fine-tune the model on a subset of the most structurally significant bars $\mathcal{P}' \subset \mathcal{P}$. We dynamically select the 5--10 most important bars based on a heuristic $\mathcal{H}$ that identifies pivot points, such as changes in tempo, time and key signatures, instrumentation, or note density. Specifically, $\mathcal{H}$ identifies candidate bars that exhibit the largest absolute changes in musical attributes relative to their preceding bars. These candidates are then ranked using a weighted scoring system that periodically alternates priority between rhythmic, harmonic, and timbral features via randomized weighting profiles. The goal of this ranking is to present the model with dynamic structural pivots during the training stage.

\textbf{LLM Guided Generation:} During inference, \textit{Text2Score} operates as a two-stage pipeline. An LLM translates a user's prompt $S$ into the structured plan $P$, which is fed into the execution-stage model. This decomposition leverages LLM reasoning for structural planning while the hierarchical decoder handles music generation.

\section{Evaluation Framework}
A key contribution of this work is the introduction of an evaluation framework designed to quantify the technical quality of generated sheet music. We propose objective metrics categorized into playability, readability, instrument utilization, and metadata adherence.

\subsection{Playability Metrics}
Playability assessment focuses on the physical constraints of human performance. We define a violation-based scoring system where $100\%$ indicates perfect adherence to instrument-specific constraints.
\begin{itemize}
    \item \textbf{Pitch Range:} For an instrument with defined MIDI range $[L, U]$, we measure the ratio of notes $n$ such that $L \le \text{pitch}(n) \le U$. Notes outside this range would be physically infeasible to play.
    \item \textbf{Pitch Span:} To account for human hand span limits, for any chord $C$, we enforce $\max(\text{pitches} \in C) - \min(\text{pitches} \in C) \le S_{max}$, where $S_{max}$ is the maximum feasible interval for the instrument (e.g., 15 semitones for piano).
    \item \textbf{Monophonic Correctness:} Certain instruments (e.g., flute, trumpet) are monophonic. For  monophonic parts, we calculate the percentage of time-steps containing only a single pitch event as opposed to chord based events.
    \item \textbf{Rhythmic Overlap:} In monophonic streams, a note onset $O_n$ shouldn't occur before the preceding note offset $E_{n-1}$. We calculate the percentage of notes that avoid this unintentional polyphony.
\end{itemize}

\textbf{Total Playability} is the macro-average of all  playability scores across all $N$ active instruments, weighted equally.

\subsection{Readability Metrics}
Readability evaluates the clarity of the symbolic encoding for rendering musical notation (engraving).
\begin{itemize}
    \item \textbf{Rhythmic Jitter:} This metric identifies quantization noise that often occurs in symbolic generative models. We flag any note with a duration of a $64^{th}$ note or shorter ($d \le 0.0625$ quarter notes), as well as any note whose onset fails to align precisely with the $64^{th}$-note metrical grid. High jitter reflects imprecise rhythm via micro-beats that clutters the score and hinders performance.
    
    \item \textbf{Rhythmic Complexity:} This metric focuses on the ease of reading rhythmic groupings by identifying excessive or unnecessary ties. We measure the ratio of tied notes to the total note count. While ties are structurally necessary, an abnormally high ratio often indicates poor beat-grouping logic that obscures the underlying meter.
    
    \item \textbf{Accidental Consistency:} Assesses the tonal coherence by measuring the percentage of notes that belong to the diatonic scale of the requested key signature. A low consistency score suggests that the model generates accidentals that do not align with the tonal centre of the piece.
    
    \item \textbf{Enharmonic Directionality:} Evaluates if accidentals are spelled logically given the key. For example, in sharp-based keys, the presence of flats is flagged as a violation. 
\end{itemize}

\textbf{Total Readability} is computed analogously to Total Playability as the macro-average of all constituent readability metric scores across all $N$ active instruments. 
We emphasize that these metrics be treated as diagnostic proxies and not as validated measures of musical quality. Rhythmic jitter can penalize legitimate tuplets or irregular subdivisions, accidental consistency can penalize deliberate chromaticism or modulation, and enharmonic directionality is key- and context-dependent. Therefore, low scores should be read as flags for inspection and not definitive judgements of quality.

\subsection{Instrumental Utilization}
To detect abandoned tracks common from multi-track models, we implement two metrics focused on temporal presence.

\begin{itemize}
    \item \textbf{Coverage Ratio:} Identifies whether an instrument remains active throughout its intended duration. It captures cases where a model begins a part but ``forgets'' to continue generating for that instrument after several bars. It is calculated as the distance between the first bar containing a note event ($m_{first}$) and the last bar containing a note event ($m_{last}$), normalized by the total bars $M_{total}$: 
    \begin{equation}
        \text{Coverage} = \frac{m_{last} - m_{first} + 1}{M_{total}}
    \end{equation}
    A low score indicates the instrument dropped out prematurely, whereas a high score indicates it persisted throughout the composition.

    \item \textbf{Active Density:} Evaluates the frequency of an instrument's participation all throughout the piece to provide a granular view of how often an instrument is actually playing. It is defined as the number of bars that contain at least one note event ($|M_{active}|$) for an instrument divided by the total number of bars in the piece ($|M_{total}|$):
    \begin{equation}
        \text{Density} = \frac{|M_{active}|}{|M_{total}|}
    \end{equation}
    Compared to Coverage Ratio, it reveals if an instrument is consistently active or only appears for isolated events.
\end{itemize}
As human-composed music rarely sounds every instrument in every bar, we treat both of these metrics as descriptive checks rather than quality scores. Coverage or density below 0.5 flags abandoned parts and warrants inspection, while higher values are not ranked.

\subsection{Prompt and Metadata Adherence}

Following \cite{bhandari2025text2midi,roy2025text2midi}, we evaluate tempo, key, and time signature matching, Cosiatec \cite{meredith2013cosiatec} based structural complexity, and semantic alignment. We replace CLAP \cite{wu2023large} with CLAMP3 \cite{wu2025clamp3universalmusic}, which operates directly on symbolic representations without audio synthesis. We additionally introduce \textbf{Instrument Match} via an LLM-as-a-judge to handle non-standardized MusicXML instrument names (e.g., "Violoncello" vs "Cello"). Furthermore, key matching accepts relative major/minor equivalents to account for limitations of key detection with Music21 library \cite{cuthbert2010music21}.


\section{Experiments}

\subsection{Implementation Details}
\textit{Text2Score} uses ModernBERT-base \cite{warner2025smarter} as the plan encoder, paired with a hierarchical decoder built on GPT-2, comprising 20 patch-level layers and 6 character-level layers with a hidden size of 768 and a maximum patch length of 2048 tokens with a patch size of 16.

We pre-train the model with AdamW optimizer at a learning rate of $1e-4$ for 30 epochs across 4 NVIDIA A100 GPUs with a batch size of 8 and 2 gradient accumulation steps. Structural fine-tuning follows for an additional 25 epochs with a learning rate of $1e-5$, batch size of 8 and 4 gradient accumulation steps, yielding an effective batch size of 32. We select the best checkpoint based on lowest validation loss.

To generate structural plans, we use GPT-5.1 as the LLM orchestrator with a 1-shot prompting strategy; we provide it with a single example of a natural language prompt with its corresponding plan, together with instructions specifying the required formatting and schema. 

\subsection{Dataset Curation}

We curated a large-scale symbolic music dataset comprising 621,396 pieces in ABC notation, compiled from MIDI-to-ABC (focusing only on quantized MIDI formats to ensure high-quality training signals), direct XML-to-ABC conversions as well as from publicly available online sources. The curated data summarised in Table \ref{Dataset-Table} covers a wide range of pieces including chamber music, symphonies, film soundtracks, folk tunes, choral and solo instrument works with a diverse mix of instruments.

\begin{table}[ht]
\centering
\begin{tabular}{lr}
\hline
\textbf{Source} & \textbf{Count} \\ \hline
ABC Notation Data & 316,118 \\
PDMX Dataset \cite{long2025pdmx} & 253,339 \\
SymphonyNet \cite{liu2022symphony} & 45,629 \\
Wikifonia Dataset \cite{simonetta2018symbolic} & 6,076 \\
ASAP Dataset \cite{foscarin2020asap} & 234 \\ \hline
\textbf{Total} & \textbf{621,396} \\ \hline
\end{tabular}
\caption{Distribution of the curated ABC notation dataset.}
\label{Dataset-Table}
\end{table}

\subsection{Evaluation Prompt Suite}
To benchmark \textit{Text2Score}, we constructed a suite of 238 prompts targeting genres with high structural and notation demands: Western classical (solo keyboard, choral, chamber and orchestral works), jazz, and cinematic scores. Prompts vary in specificity, ranging from explicit constraints on instrumentation, key, and time signature to high-level descriptive instructions covering emotional arcs and structural development (e.g., ``start with a solo cello theme and gradually build to a full orchestral climax in the middle section''). We conducted both the objective evaluation and the subjective listener study on this prompt set.

\subsection{Baselines}
\label{baselines}
To evaluate the efficacy of our sub-task decomposition approach, we benchmark \textit{Text2Score} against models at two ends of the generative spectrum. For zero-shot agentic composition, we compare against ComposerX \cite{deng2024composerx}, a multi-agent system powered by LLMs (we use GPT-5.1 for a fair comparison) that composes ABC notated symbolic music without fine-tuning. For end-to-end neural generation, we compare against Text2Midi-InferAlign \cite{roy2025text2midi}, MIDI-LLM \cite{wu2025midi} and MIDILM \cite{li2026midilm}, which represent the current state-of-the-art in direct text-to-token generation of MIDI music. This selection allows us to assess whether our hybrid approach can overcome the structural simplifications common in pure LLM-based composition and the alignment issues of end-to-end models.

\subsection{Subjective Evaluation}
We invited 24 expert musicians to evaluate outputs from the three generative frameworks described in the Baselines Section: ComposerX, Midi-LLM, and \textit{Text2Score}. Notably, 119 of 238 (50\%) evaluation prompts yielded invalid XML outputs under ComposerX, a limitation consistent with the authors' own observation that musical elements are sometimes inadequately translated into ABC notation by the musician agents \cite{deng2024composerx}. All prompts selected for the study were therefore drawn from ComposerX's valid outputs, giving this baseline the most favourable conditions for comparison. Using the Goldsmiths Musical Sophistication Index\footnote{\url{https://www.gold.ac.uk/music-mind-brain/gold-msi/}} as a self-report measure, 14 participants (58\%) reported more than 10 years of experience in instrumental or vocal practice, formal music theory, or performance training; the remaining reported fewer than 10 years.

For each music example, we rendered the output of each model as a synchronised video of the MuseScore notation alongside high-quality synthesized audio using MuseScore core sounds. To avoid listener fatigue, each participant was randomly assigned 2 prompts from the full set of variants, yielding 6 videos from the 3 models in random order. Participants rated each score on the following criteria, each accompanied by a detailed description to minimise ambiguity:

\begin{enumerate}
    \item \textbf{Prompt Adherence:} How accurately does the generated music reflect the constraints of the text prompt?
    \item \textbf{Readability \& Engraving:} How clear and standard is the musical notation for a performing musician? 
    \item \textbf{Musicality \& Expressive Intent:} How aesthetically pleasing and musically expressive is the composition? 
    \item \textbf{Authenticity to Professional Composition:} How closely does the generated score resemble the work of a professional human composer? 
    \item \textbf{Usability for Professional Composition:} To what extent could this score serve as a viable foundation for a professional composer requiring only minimal edits?
\end{enumerate}


\section{Results and Analysis}

\begin{table}[th]
\centering
\resizebox{\columnwidth}{!}{%
\begin{tabular}{@{}lccccc@{}}
\toprule
\textbf{Metric} & \textbf{Text2Score} & \textbf{ComposerX} & \textbf{Midi-LLM} & \textbf{Infer-Align} & \textbf{MidiLM} \\ \midrule

\multicolumn{6}{c}{\textit{Generation Efficiency}} \\ \midrule
Valid Files Gen. & 97.48\% & 50.00\% & \textbf{100.00}\% & 99.58\% & 97.90\% \\
Total API Cost & \textbf{\$2.18} & \$91.56 & - & - & - \\
Total API Calls & \textbf{238} & 4,484 & - & - & - \\ \midrule

\multicolumn{6}{c}{\textit{Playability}} \\ \midrule
Pitch Range & \textbf{97.75}\% & 85.11\% *** & 95.77\% ** & 96.34\% * & 93.73\% *** \\
Monophonic Corr. & \textbf{97.49}\% & 91.00\% *** & 69.96\% *** & 89.43\% *** & 86.74\% *** \\
Pitch Span & \textbf{99.56}\% & 98.74\% * & 95.31\% *** & 97.33\% *** & 97.44\% *** \\
Rhythmic Overlap & \textbf{99.24}\% & 90.06\% *** & 92.81\% *** & 98.35\% *** & 95.48\% *** \\
\textbf{Total Playability} & \textbf{98.81}\% & 90.29\% *** & 90.31\% *** & 95.65\% *** & 93.93\% *** \\ \midrule

\multicolumn{6}{c}{\textit{Readability}} \\ \midrule
Rhythmic Comp. & 97.28\% & \textbf{98.51}\% * & 79.83\% *** & 89.46\% *** & 82.56\% *** \\
Rhythmic Jitter & \textbf{96.69}\% & 83.30\% *** & 62.73\% *** & 95.15\% * & 87.36\% *** \\
Accidental Consist. & 92.38\% & \textbf{98.40}\% *** & 84.82\% *** & 82.36\% *** & 84.88\% *** \\
Enharmonic Dir. & 99.40\% & \textbf{99.54}\% & 89.08\% *** & 92.47\% *** & 80.80\% * \\
\textbf{Total Readability} & \textbf{96.19}\% & 95.22\% * & 79.04\% *** & 90.10\% *** & 84.47\% *** \\ \midrule

\multicolumn{6}{c}{\textit{Prompt Adherence}} \\ \midrule
Tempo Match & 92.24\% & 99.16\% *** & \textbf{100.00}\% *** & 45.34\% *** & 90.56\% \\
Key Sig. Match & 97.41\% & \textbf{100.00}\% * & 40.00\% *** & 25.42\% *** & 56.65\% *** \\
Time Sig. Match & 99.14\% & \textbf{100.00}\% & 48.70\% *** & 47.46\% *** & 97.42\% \\
Instrument Match & \textbf{83.15}\% & 55.07\% *** & 50.46\% *** & 23.82\% *** & 41.39\% *** \\
CLAMP3 Sim. & \textbf{0.1446} & 0.1266 ** & 0.0825 *** & 0.0207 *** & 0.0935 *** \\ \midrule

\multicolumn{5}{c}{\textit{Structural Complexity \& Instrument Utilization}} \\ \midrule
Structure & \textbf{3.18} & 2.34 *** & 2.40 *** & 2.30 *** & 2.09 ***\\
Coverage Ratio † & 79.53\% & 90.27\% & 84.24\% & 38.79\% & 49.19\% \\
Active Density † & 76.13\% & 88.55\% & 81.01\% & 36.69\% & 44.37\% \\ \bottomrule
\end{tabular}%
}
\caption{Comprehensive objective evaluation across generations, playability, readability, adherence, and structure. Metrics were calculated exclusively on valid files. Significance levels ($^{*}\, p < 0.05$, $^{**}\, p < 0.01$, $^{***}\, p < 0.001$) indicate Welch's t-test comparisons against Text2Score. † Diagnostic checks; significance is not reported.}
\label{tab:master_results}
\end{table}


\begin{table}[th]
\centering
\resizebox{\columnwidth}{!}{%
\begin{tabular}{@{}lccccc@{}}
\toprule
\multirow{2}{*}{\textbf{Dimension}} & \textbf{Text2Score} & \multicolumn{2}{c}{\textbf{ComposerX}} & \multicolumn{2}{c}{\textbf{Midi-LLM}} \\ \cmidrule(l){3-4} \cmidrule(l){5-6}
 & \textbf{Mean} & \textbf{Mean} & \textbf{\textit{p}-value} & \textbf{Mean} & \textbf{\textit{p}-value} \\ \midrule
Prompt Adherence & \textbf{3.48} & 2.94 & $0.003$ & 1.67 & $<0.001$ \\
Readability & \textbf{3.98} & 2.92 & $<0.001$ & 1.79 & $<0.001$ \\
Musicality & \textbf{3.52} & 2.92 & $<0.001$ & 1.69 & $<0.001$ \\
Authenticity & \textbf{3.12} & 2.44 & $<0.001$ & 1.44 & $<0.001$ \\
Usability & \textbf{3.44} & 2.65 & $<0.001$ & 1.52 & $<0.001$ \\
\bottomrule
\end{tabular}%
}
\caption{Subjective evaluation results (Mean Opinion Scores on a 5-point scale). Significance ($p$-values) calculated using a Linear Mixed-Effects Model (LMM) accounting for prompt and participant random effects.}
\label{tab:subjective_results}
\end{table}

\begin{table}[H]
\centering
\resizebox{\columnwidth}{!}{%
\begin{tabular}{@{}lcccc@{}}
\toprule
\textbf{Metric} &
\shortstack{\textbf{Gemma4}\\\textbf{31B-IT}} &
\shortstack{\textbf{Qwen2.5}\\\textbf{14B-IT}} &
\shortstack{\textbf{Gemma4}\\\textbf{12B-IT}} &
\shortstack{\textbf{GLM4.1V}\\\textbf{9B}} \\
\midrule

Valid Files Gen. & \textbf{98.32\%} & \textbf{98.32\%} & \textbf{98.32\%} & 97.06\% \\ \midrule

\multicolumn{5}{c}{\textit{Playability}} \\ \midrule
Pitch Range & 97.18\% & \textbf{97.93\%} & 95.31\% ** & 96.14\% \\
Monophonic Corr. & \textbf{98.49\%} & 98.00\% & 94.57\% *** & 95.00\% *** \\
Pitch Span & \textbf{99.57\%} & 99.51\% & 98.76\% ** & 98.81\% * \\
Rhythmic Overlap & \textbf{99.79\%} & 99.67\% & 99.36\% ** & 99.59\% \\
\textbf{Total Playability} & \textbf{98.58\%} & 98.26\% & 97.14\% *** & 97.48\% ** \\ \midrule

\multicolumn{5}{c}{\textit{Readability}} \\ \midrule
Rhythmic Comp. & 95.50\% & 96.10\% & \textbf{96.26\%} & 94.90\% \\
Rhythmic Jitter & 97.04\% & \textbf{98.75\%} *** & 96.79\% & 98.10\% * \\
Accidental Consist. & \textbf{91.73\%} & 88.99\% *** & 86.07\% *** & 82.39\% *** \\
Enharmonic Dir. & \textbf{98.94\%} & 98.58\% & 98.37\% * & 98.29\% * \\
\textbf{Total Readability} & \textbf{95.84\%} & 94.91\% & 93.48\% *** & 92.21\% *** \\ \midrule

\multicolumn{5}{c}{\textit{Prompt Adherence}} \\ \midrule
Tempo Match & \textbf{95.73\%} & 94.44\% & 92.31\% & 92.21\% \\
Key Sig. Match & \textbf{88.89\%} & 79.91\% ** & 80.34\% * & 69.26\% *** \\
Time Sig. Match & 99.57\% & \textbf{100.00\%} & \textbf{100.00\%} & \textbf{100.00\%} \\
Instrument Match & \textbf{87.43\%} & 77.03\% *** & 78.87\% *** & 73.26\% ***\\
CLAMP3 Sim. & \textbf{0.1406} & 0.0967 *** & 0.1166 *** & 0.1169 ***\\ \midrule

\multicolumn{5}{c}{\textit{Structural Complexity \& Instrument Utilization}} \\ \midrule
Structure & 3.20 & 3.64 * & 3.49 * & 3.45 \\
Coverage Ratio † & 78.60\% & 72.83\% & 76.69\% & 85.38\% \\
Active Density † & 74.82\% & 70.25\% & 72.61\% & 81.46\% \\ \bottomrule
\end{tabular}%
}
\caption{Ablated objective evaluation results highlighting the sensitivity of Text2Score when paired with various open-source LLM orchestrators of different sizes. Significance levels ($^{*}\, p < 0.05$, $^{**}\, p < 0.01$, $^{***}\, p < 0.001$) indicate Welch's independent t-test comparisons against the Gemma4 31B-IT. † Diagnostic checks; significance is not reported.}
\label{tab:ablation_results}
\end{table}

\subsection{Objective Evaluation Analysis}

\textbf{Generation Efficiency:} As shown in Table \ref{tab:master_results}, ComposerX yields valid outputs for only 50.00\% of prompts, largely due to its inability to consistently compile valid ABC notation, producing mismatched bar durations or missing part declarations. Its reliance on multiple LLM calls also incurs significant cost: 4,484 calls totalling \$91.56 versus \textit{Text2Score's} \$2.18 for all 238 prompts.

\textbf{Playability:} \textit{Text2Score} achieves the strongest overall playability (98.81\%), suggesting that LLMs possess a capable understanding of physical instrumental constraints when tasked with defining them explicitly in a planning stage. However, this latent knowledge is often lost when an LLM outputs symbolic music directly without fine-tuning, consistent with observations in the ComposerX study. End-to-end baselines (Midi-LLM, Infer-Align and MidiLM) exhibit difficulties with monophonic instruments (e.g., flutes, trumpets).

\textbf{Readability:} ComposerX achieves high scores across several individual readability metrics. However, this may be indicative of structurally simplistic outputs. For instance, professional compositions may not exhibit accidental consistencies as high due to intentional modulations and changes in tonal colour. ComposerX also struggles with rhythmic jitter (83.30\%), indicating difficulties with metrical grid placement and inter-voice alignment. This limitation is also noted in their original study. The end-to-end models underperform across all readability metrics.

\textbf{Prompt Adherence:} ComposerX interprets text well, matching key and time signatures reliably, but its adherence is undermined by instrument name hallucinations and failures to instantiate valid parts in multi-instrument prompts. \textit{Text2Score} achieves the highest CLAMP3 text-music similarity (0.1446). Its slightly lower tempo match relative to Midi-LLM is partly attributable to a formatting translation gap during generation (e.g., a requested tempo of 75 BPM for eighth notes rendered as 150 BPM for quarter notes), rather than a musical misinterpretation. The end-to-end baselines struggle across most adherence metrics.

\textbf{Instrument Utilization and Structural Complexity:} Utilization is read as a descriptive check and not a ranking. \textit{Text2Score}, ComposerX and Midi-LLM all sit above our flagging floor (76–90\% coverage), indicating that parts are instantiated and sustained; ComposerX's near-ceiling values are consistent with a simplistic timbral approach in which every instrument plays in almost every bar. The check does its work on Infer-Align and MidiLM, whose low coverage (38.79\% and 49.19\%) signals abandoned tracks. \textit{Text2Score} balances reasonable instrumental participation with the highest structural complexity (3.18), which may suggest more varied and well-developed musical textures.

\begin{figure*}[th]
\centering
\includegraphics[width=.95\textwidth]{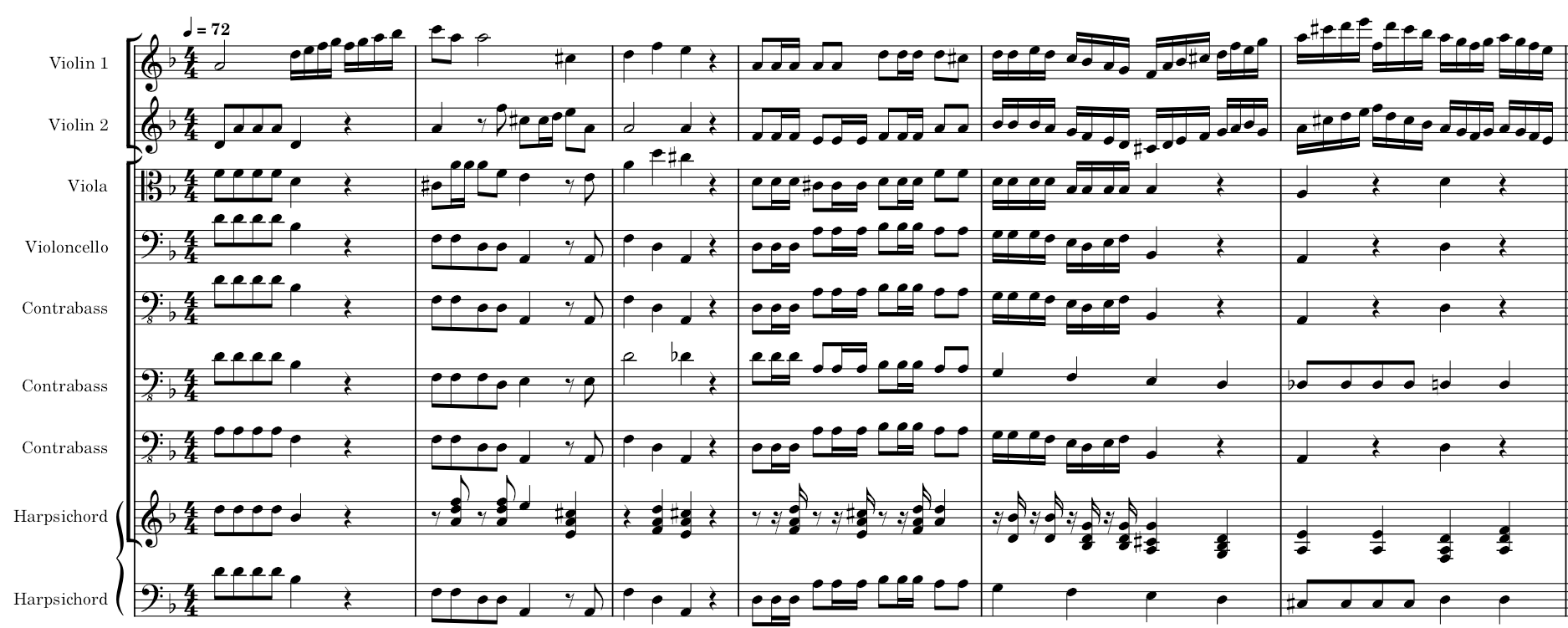}
\caption{An illustrated example of a \textit{Text2Score} generated score using the following caption ``\textit{A baroque-style double concerto for two violins and a string orchestra in D minor. The tempo is a steady 4/4 at 72 BPM. The orchestra should provide a solid continuo-style accompaniment with the cellos and double basses playing a consistent bass line. The music should feel formal and academic, but with a lot of emotional depth in the minor key. Make sure the parts are well-balanced so the soloists can always be heard over the ensemble. Let the tension build through the use of suspensions and resolutions.}'' In this example, we see Text2Score generates the score in D melodic minor, consistent with instructions - ``lot of emotional depth in the minor key."}
\label{fig:score}
\end{figure*}

\subsection{Subjective Evaluation Analysis}

As shown in Table \ref{tab:subjective_results}, \textit{Text2Score} consistently outperforms both baselines across all five dimensions, with all improvements reaching statistical significance ($p < 0.001$). The most pronounced advantage is in Readability (3.98), where \textit{Text2Score} scores above ComposerX (2.92) and Midi-LLM (1.79), corroborating the objective engraving metrics. This may also stem from richer engraving detail in the generated scores, including dynamics, tempo markings, articulations, and accidentals. \textit{Text2Score} also leads in Musicality (3.52) combined with Usability (3.44), indicating that relative to the baselines, expert musicians found its outputs the most aesthetically pleasing with great potential as a foundation for further composition. Midi-LLM performs poorly across the board, reflecting the fundamental unsuitability of MIDI-based end-to-end generation for notation quality. While ComposerX remains competitive in Musicality (2.92) and Prompt Adherence (2.94), its lower scores in Authenticity (2.44) and Usability (2.65) suggest its outputs lack the depth and structural coherence expected of professional scores. We provide an example of a \textit{Text2Score} generated output in Figure \ref{fig:score}.

\subsection{Plan Quality Ablation}

To isolate the contribution of the plan, we degrade plan quality by replacing GPT-5.1 with four open-source LLMs spanning 9B–31B parameters (Table \ref{tab:ablation_results}), the smaller of which produce internally inconsistent plans. Playability remains stable across orchestrators (97.14–98.58\%), confirming that physical performance is largely decoupled from the planning LLM and governed by the execution model. Readability and prompt adherence, by contrast, scale with orchestrator capability: Total Readability falls from 95.84\% (Gemma4 31B) to 92.21\% (GLM4.1V 9B), driven largely by accidental consistency, and this same weakness surfaces in key signature matching, which degrades sharply from 88.89\% to 69.26\%, as smaller models specify tonal attributes less reliably in the plan. Instrument matching follows the same ordering, dropping from 87.43\% to 73.26\%. Higher Cosiatec compression scores among the smaller orchestrators (3.45–3.64, vs. 3.18–3.20 for GPT-5.1 and Gemma4 31B) may partly reflect a greater reliance on ostinatos (repeated patterns) rather than motivic development. Output quality therefore tracks plan quality directly, indicating that generation is governed by the plan rather than the decoder's own priors, while performability remains stable regardless of orchestrator scale.


\subsection{Limitations and Future Work}

A potential failure mode of \textit{Text2Score} may arise if the LLM-generated inference plan diverges substantially from plans seen during training. While structural fine-tuning mitigates large discrepancies by exposing the model to sparse, non-consecutive bar plans, pronounced semantic mismatches can still cause the model to deviate from the user's intended prompt. This can be partially addressed through careful prompt engineering or, thanks to the transparency of our two-stage design, by a human composer directly inspecting and overriding the plan before generation. Looking ahead, the planning stage could be further enriched through retrieval-augmented generation over curated musical knowledge bases, enabling richer compositional control.

A further limitation lies in the expressive resolution of the plan. While the bar-wise plan captures a structural skeleton with some expressive attributes such as dynamics, more granular details conveyed in a prompt such as specific harmonic textures or voice-leading instructions are not explicitly represented. Future work could address this by leveraging our open-sourced dataset to develop richer annotations that combine the structural outline with textual descriptors to capture these finer musical details.

\section{Conclusion}

We presented \textit{Text2Score}, a two-stage framework for generating sheet music from natural language prompts that offers an alternative training paradigm in the absence of large-scale aligned text-music pairs. By decomposing generation into an explicit planning stage followed by a dedicated execution stage, we demonstrate that separating musical reasoning from note-level generation yields substantial gains over end-to-end approaches and pure LLM-based agentic composition. We further contribute a suite of objective metrics including readability and playability metrics alongside a subjective evaluation, across which \textit{Text2Score} achieves statistically significant improvements. We hope this work encourages further exploration at the intersection of LLMs and symbolic music representations, where the reasoning capabilities and musical knowledge of modern LLMs remain largely untapped.

\section*{Acknowledgments}

This work was supported by UKRI and EPSRC (grant EP/S022694/1) and by SUTD’s Kickstart Initiative (grant number SKI 2021 04 06) and MOE (grant number MOE-T2EP20124-0014). Additionally, the authors acknowledge support from the IEEE Signal Processing Society under the Signal Processing Society Scholarship Program. 

\bibliography{aaai2027}


\end{document}